# LLM-Assisted Authentication and Fraud Detection

Emunah S. S. Chan, Aldar C-F. Chan

**Abstract**

User authentication and fraud detection face growing challenges as digital systems expand and adversaries adopt increasingly sophisticated tactics. Traditional knowledge-based authentication remains rigid, requiring exact string matches that fail to accommodate natural human memory and linguistic variation. Meanwhile, fraud-detection pipelines struggle to keep pace with rapidly evolving scam behaviours, leading to high false-positive rates and frequent retraining cycles. This research introduces two complementary LLM-enabled solutions: (1) an LLM-assisted authentication mechanism that evaluates semantic correctness rather than exact wording, supported by document segmentation and a hybrid scoring method combining LLM judgment with cosine-similarity metrics; and (2) a RAG-based fraud-detection pipeline that grounds LLM reasoning in curated evidence to reduce hallucinations and adapt to emerging scam patterns without model retraining. Experiments show that the authentication system accepts 99.5% of legitimate non-exact answers while maintaining a 0.1% false-acceptance rate, and that the RAG-enhanced fraud detector reduces false positives from 17.2% to 3.5%. Together, these findings demonstrate that LLMs can significantly improve both usability and robustness in security workflows, offering a more adaptive, explainable, and human-aligned approach to authentication and fraud detection.

## 1. Introduction

Modern digital systems face a persistent tension between security and usability. Authentication mechanisms must be strong enough to prevent unauthorized access yet flexible enough to accommodate natural human variability. However, contemporary systems continue to rely predominantly on “what you know” factors — passwords, PINs, and security questions — which are inherently rigid. These mechanisms require exact string matches, leaving no allowance for paraphrasing, partial recall, or minor linguistic variation. This strictness stands in contrast to the approximate and contextual nature of human memory and results in frequent authentication failures: a single misplaced character can lock out a legitimate user, and password-reset workflows similarly demand precise answers to security questions. Such rigidity increases operational support costs and disproportionately affects users with cognitive impairments, memory challenges, or language differences. Meanwhile, many security questions depend on information that can be publicly discovered or socially engineered, further eroding their reliability.

Simultaneously, online fraud has expanded in scale and sophistication. Global financial losses continue to grow, fueled by phishing schemes, investment scams, romance scams, and business-email-compromise attacks. INTERPOL's *Global Financial Fraud Assessment* (May 2024) reports that financial fraud has reached epidemic levels and describes financial fraud as a "massive and global" problem driven by the rapid expansion of organized cyber-enabled crime [22]. Fraudsters rapidly evolve their tactics, often exploiting AI-generated content and impersonation strategies to deceive victims. Traditional machine-learning-based fraud-detection systems struggle to keep pace because they depend on static classifiers that must be constantly retrained as the fraudsters' tactics evolve.

These challenges underscore the need for authentication and fraud-detection systems that can adapt to dynamic linguistic behaviour and reason about human communication rather than rely on deterministic pattern matching. Large language models (LLMs) offer a promising alternative for both problems. Their ability to interpret natural language, assess semantic similarity, and reason over context makes them well suited for tasks involving human communication. This research examines how LLMs can be integrated into authentication and fraud-detection workflows to create user-authentication systems that are more intuitive and user-friendly, and fraud-detection systems that are more adaptive and resilient to the evolving tactics of fraudsters.

This research introduces two LLM-based solutions:

1. LLM-assisted user authentication: A semantic evaluation mechanism that accepts natural-language answers without requiring exact matches.
2. LLM-assisted fraud detection using RAG (Retrieval Augmented Generation): A context-aware pipeline that retrieves relevant information (confirmed scam message samples, corporate policies) before generating a risk score.

Both approaches leverage LLMs' strengths in natural-language understanding and contextual reasoning and address the limitations of LLMs.

## 1.1 LLM-assisted User Authentication

In the LLM-assisted user authentication scheme, the system supplies an LLM with a user-specific profile document — such as personal information or recent account activity. Rather than relying on manually crafted security questions, the LLM automatically generates personalized challenge questions along with corresponding reference answers. During login, the user responds in natural language, and the LLM evaluates these responses based on semantic meaning rather than exact character-level matching. This enables correct but non-exact answers — paraphrases, reordered wording, or additional contextual details — to be accepted, greatly improving user friendliness.

To overcome inherent limitations of LLMs, the design introduces two key innovations. First, the user profile document is segmented before being passed to the LLM, ensuring that generated questions draw evenly from all parts of the document instead of clustering at the beginning or end, a common tendency of LLMs. Second, the LLM's evaluation is combined with a statistical similarity metric (e.g., cosine similarity) computed between the embeddings of the user's answer and the reference answer. This hybrid scoring mechanism enhances robustness: it allows semantically close responses to be accepted even if the LLM initially flags them, while still reducing the chance of mistakenly accepting incorrect answers. When similarity thresholds are set more leniently for user friendliness, the system compensates by adaptively increasing the number of questions asked, thereby maintaining overall security.

Experimental results with Llama-3.3 demonstrate the feasibility of the approach: the LLM correctly accepts 99.5% of valid non-exact answers while maintaining a very low false-acceptance rate of 0.1%.

Overall, this design introduces a new authentication factor — textual, resettable, adaptive, and nondeterministic — enabled by LLM-based semantic reasoning. It transforms authentication from a rigid memory test into a natural, conversational interaction. For elderly users who often struggle with remembering passwords or precise login credentials, this approach offers a more forgiving, less stressful, and even enjoyable experience while preserving strong security assurance. By reframing authentication as natural language dialogue, the method allows users to respond in their own words without compromising trustworthiness.

## 1.2 Adaptive Scam Detection using LLMs

The proposed LLM-enabled fraud and scam detection system introduces an adaptive, context-aware method for identifying fraudulent messages by combining LLMs with RAG. Traditional fraud-detection techniques rely on static classifiers that match fixed patterns or hand-crafted features. Because scammers constantly change wording, structure, and psychological tactics, these models rapidly become outdated and require repeated retraining. In contrast, the proposed system leverages the linguistic reasoning capabilities of LLMs to interpret messages semantically, enabling detection that aligns with how humans recognize suspicious behaviour.

The system operates through a multi-stage pipeline. When a message is received, the LLM first extracts its key features — intent, urgency, tone, requested actions, entities mentioned, and contextual clues. These extracted elements are then used to retrieve relevant evidence from multiple external repositories, such as confirmed scam databases, organisational policy documents, and domain-specific reference materials. This retrieved information is incorporated into a structured prompt, which is returned to the LLM for final assessment. Based on the combined context, the LLM produces both a scam-likelihood score and an explanation that supports transparency and trust.

This design addresses several limitations of standalone LLMs. LLMs used in isolation tend to generate high false-positive and false negative rates, partly because they interpret text without grounding in verified evidence. By integrating RAG, the system ties the LLM's reasoning to authoritative sources, drastically reducing over-sensitivity. Moreover, because updates occur at the database level, the system remains adaptive to evolving scam tactics without requiring any LLM retraining. The structured prompting process also mitigates hallucinations by anchoring the model's reasoning in retrieved facts.

The key innovations of this approach include its grounding of LLM reasoning in dynamic evidence, its adaptability without retraining, its explainable output, and its robustness against rapidly evolving fraud patterns. Together, these features create a flexible, scalable, and resilient fraud-detection framework that surpasses traditional classifier-based methods.

### 1.3 Contributions

The contributions of this work are three-fold. First, it introduces a new semantic knowledge-based authentication factor that replaces rigid word-for-word string matching with LLM-driven natural-language evaluation. Through document segmentation and a hybrid scoring method combining LLM judgment with cosine similarity, the system overcomes positional bias and enables flexible, human-like verification. Second, it presents a RAG-grounded fraud-detection pipeline that anchors LLM reasoning in curated scam corpora and organisational policies, reducing hallucinations and allowing rapid adaptation to emerging scam tactics without retraining. Third, it empirically demonstrates performance improvements: the authentication system accepts 99.5% of legitimate non-exact answers while keeping false acceptances at 0.1%, and the RAG-enhanced fraud detector reduces false positives to below 3%. Together, these innovations show that LLM-enabled semantic reasoning can simultaneously improve usability, robustness, and explainability across critical security workflows.

## 2. Related Work

### 2.1 User Authentication

Classic multi-factor authentication (MFA) traditionally relies on three factors — what you know, what you have, and who you are [13]. Over time, however, researchers have proposed new factors that extend beyond this canonical trio. For example, Brainard et al. introduce a "fourth factor," somebody you know, by enabling peer vouching for emergency access and examining both cryptographic and social-engineering implications [4]. Location has also been explored as an authentication factor: Bartłomiejczyk et al. design a mobile-environment protocol that incorporates positional data to strengthen authentication under resource constraints, demonstrating its security and performance properties [2].

In cyber-physical settings, Chan and Zhou argue that authentication must account for both digital and physical presence. To mitigate substitution attacks in electric-vehicle charging, they propose splitting the challenge–response exchange across wireless and physical (charging-cable) channels while binding the vehicle's physical presence through an onboard component [5]. Chan et al. further extend the space of authentication factors by introducing a lightweight, leakage-resilient factor based on historical data for IoT devices, enabling scalable verification with minimal storage requirements [6]. Beyond introducing new factors, some work has focused on improving the management of existing ones. Smith et al. show that a dedicated secondary-factor manager can substantially reduce user errors and the time needed for 2FA setup or replacement, underscoring the importance of lifecycle support in real-world security [20].

Despite this proliferation of new authentication factors, the "what you know" factor — passwords, PINs, passcodes — remains the most widely used in practice. Yet it continues to suffer from long-standing usability limitations, particularly when users forget credentials and must undergo cumbersome recovery processes.

More recently, researchers have begun exploring LLM-aware authentication frameworks. Rehman et al. demonstrate that integrating LLM reasoning into IoT authentication can improve adaptivity by interpreting contextual device signals [15]. Sharma and Rani similarly apply LLMs to V2V/V2I authentication, proposing a context-sensitive scheme that reduces latency while maintaining security for autonomous-vehicle communication [18]. Notably, however, none of these works use LLMs to enhance the user-friendliness of the "what you know" factor, leaving an important gap in the literature.

The LLM-assisted authentication mechanism proposed in this work fills that gap by introducing a semantic knowledge factor, a new variant of the traditional "what you know" category. Instead of requiring exact-match answers to security questions, users respond in natural language. Authentication is granted based on the semantic equivalence between the user's response and model answers, assessed through a hybrid scoring method that combines LLM reasoning with statistical similarity. This reframes the knowledge factor to better reflect human memory and linguistic variation, moving beyond location- or social-based factors [2, 4] as well as beyond data-centric historical-factor designs [6].

Where cyber-physical schemes [5] and context-aware IoT/V2X frameworks [15, 18] adapt which signals are used for authentication, our approach instead adapts how user knowledge is evaluated, making the process more inclusive, conversational, and tolerant of natural linguistic variation while still controlling over-permissiveness through explicit statistical similarity thresholds between user and model answers.

## 2.2 Fraud and Scam Detection

The anti-scam literature spans several directions, including interaction mediation, dataset construction, agentic LLM systems, and robustness analysis. Pandit et al. propose a phone-based virtual assistant that answers unknown calls, engages callers in conversation, and determines whether the caller is a human or a robot-caller — mitigating spoofing attacks and improving user experience compared to traditional blocklists [14].

A growing body of work investigates LLMs for scam and fraud detection. Jiang presents a general pipeline for applying foundation models to phishing, romance scams, and advance-fee fraud, and highlights key integration and evaluation gaps across linguistic contexts [10]. Shen et al. further examine LLM-based phone-scam detection and identify practical obstacles, including biased datasets, low recall rates, and hallucination-prone outputs [19]. From a user-experience perspective, Alhulifi et al. compare LLM-powered and content-based phone-scam detectors on call transcripts, showing how usability and user trust influence real-world adoption [1].

At the systems level, Nakano, Koide, and Chiba develop an agentic LLM capable of autonomously gathering multi-modal evidence — webpage text, screenshots, DNS records, and user reviews — to classify scam websites with high accuracy across multiple languages and scam categories [12]. For dynamic, evolving conversations, Senol, Agrawal, and Liu propose a two-stage fraud framework: the system first flags suspicious dialogs, then applies a one-class drift detector to distinguish benign topic shifts from manipulative ones; if drift is detected, an LLM judges whether the shift signals fraudulent intent, thereby improving both accuracy and interpretability [17]. Robustness concerns are underscored by Chang et al., who construct a fine-grained dataset of adversarially perturbed scam messages and show that small, targeted edits can cause large misclassification drops in LLMs — motivating research into adversarial training and structured perturbation methods [7].

Relative to this literature, our RAG-based fraud and scam detector explicitly grounds LLM judgments in auditable, up-to-date corpora — including internal policies, curated scam exemplars, and vetted knowledge bases — prior to classification. This design directly addresses two persistent challenges: (i) hallucinations and poor calibration, which we mitigate by constraining the model to reason only over retrieved evidence; and (ii) dataset staleness and low recall, which we address by refreshing the retrieval corpus rather than retraining the model itself. In essence, our detector prioritizes explainability and adaptability, adopting the principle of refreshing the evidence, not the weights.

## 3. Design of LLM-Assisted User Authentication

### 3.1 System Architecture

The core idea behind LLM-assisted user authentication is to leverage an LLM to generate personalized challenge questions based on user-specific data stored in the target system, and to evaluate user responses semantically rather than through exact string matching. This approach transforms authentication into a conversational, human-friendly interaction, reducing cognitive load — especially for users such as older adults who may struggle to recall complex passwords.

Figure 1 illustrates the high-level architecture of the LLM-assisted authentication workflow. When a user attempts to log in, the authentication pipeline proceeds as follows:

1. The system retrieves the user's personal documents and relevant data stored in its database.
2. The system constructs a prompt instructing the LLM to generate a set of security challenge questions along with corresponding reference answers.
3. These questions are presented to the user sequentially, and the user responds in natural language.
4. The LLM evaluates the correctness of each response. It is explicitly instructed to assess semantic accuracy rather than enforce exact string matches.
5. The LLM returns a grade or confidence score for each answer.
6. The system computes the statistical similarity between the user's response and the LLM-generated reference answer, combining this similarity score with the LLM's evaluation to determine whether the login attempt should be accepted. If needed, the system may present additional security questions to further strengthen the assurance of the user's authenticity.

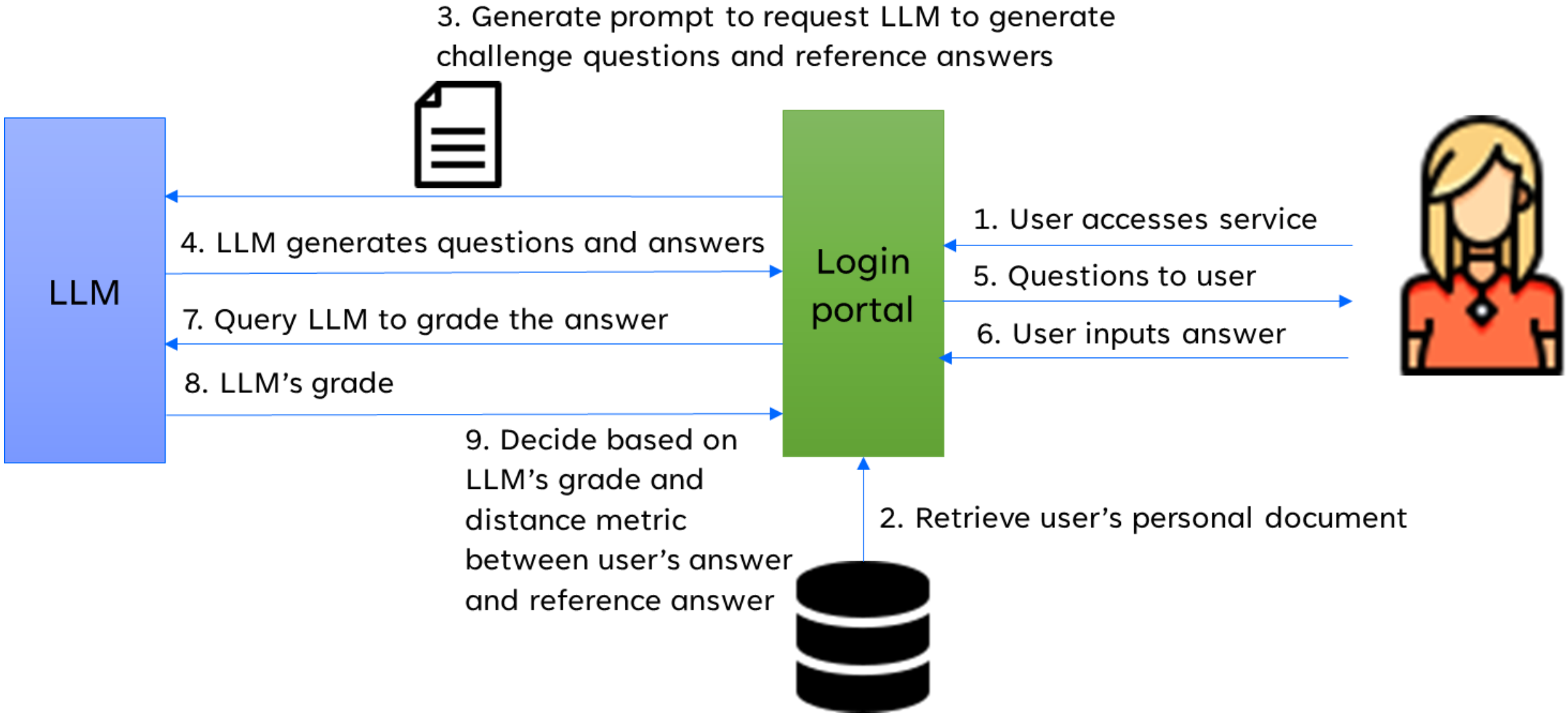

Figure 1. System Architecture of LLM-assisted User Authentication

### 3.2 Uneven Distribution of LLM-generated Questions

Prior work shows that LLMs tend to focus on the beginning and end of a text when generating summaries [23]. To examine whether a similar positional bias appears when LLMs generate security questions from user documents, we conduct experiments to test whether the models select questions unevenly across a document. This matters for security: if most questions originate from the start of a document, an adversary could more easily gather the relevant information from external sources and increase their chances of answering correctly.

To assess this risk, we evaluate ChatGPT-4 and Llama-3.3 using three different documents. Each model is asked to generate 20 security questions per document. For analysis, each document is divided into four equal word-count segments, and every generated question is assigned to the segment from which its content originates. We then compute the percentage of questions drawn from each segment and average the results across the three documents to obtain the overall distribution. As shown in Figure 2, both models display a clear positional bias, generating disproportionately more questions from the beginning and end of the documents. In short, the security questions produced by ChatGPT-4 and Llama-3.3 are unevenly distributed across the document.

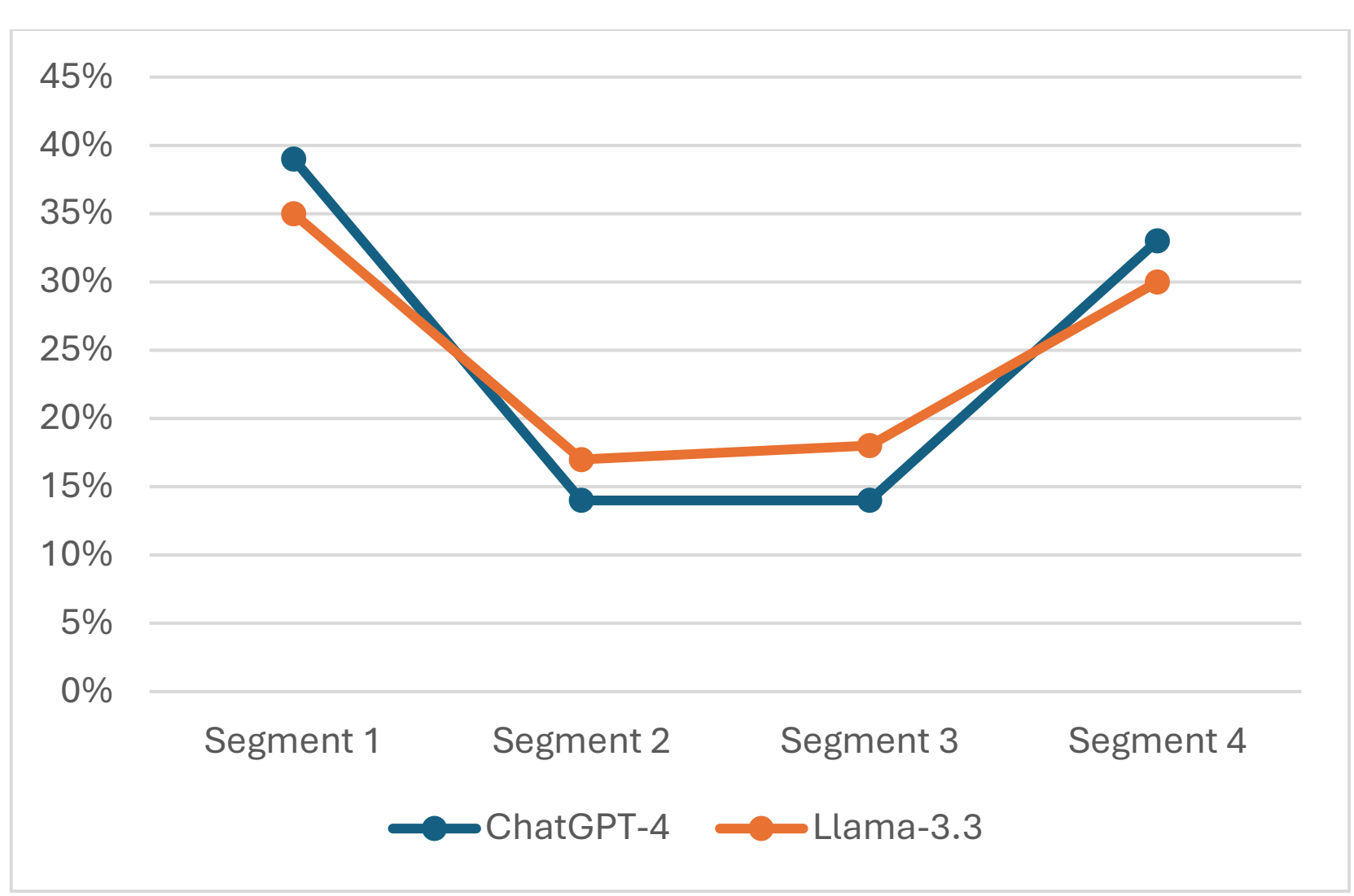

Figure 2. Average percentage of security questions from different segments of three documents generated by ChatGPT-4 and Llama-3.3

### 3.3 Detailed Implementation

Figure 3 shows how the LLM-assisted user authentication system is implemented. The proposed authentication system operates by leveraging a user's personal document as the foundation for generating individualized security questions through an LLM. When a user initiates an authentication attempt, the system first retrieves the relevant personal document and prepares it for question generation. A key step in this preparation is the segmentation of the document into multiple sections. This design choice directly addresses the tendency of LLMs: when prompted to generate questions from a text, they disproportionately select content from the beginning and the end of the document. Such positional bias reduces the diversity and unpredictability of generated questions, thereby weakening the security of the authentication process. By dividing the document into segments and requiring the LLM to generate questions from each segment, the system ensures that the resulting security questions are more evenly distributed across the entire document. This segmentation mechanism reduces predictability and thereby strengthens the overall robustness of the authentication protocol.

Once the document is segmented, the system queries the LLM to generate the same number of question–answer pairs for each segment until the required number of questions has been produced. These questions are then presented to the user, who provides responses accordingly. Each response undergoes a two-stage evaluation process. First, the LLM is asked to judge whether the user's answer is contextually correct or acceptable. Second, the system converts the user's answer and the LLM-generated model answer into vectors using sentence embedding and then computes the cosine similarity between the two vectors. [1] This similarity score is compared against a predefined threshold to determine whether the answer is sufficiently close to the expected response.

The cosine-similarity threshold plays a central role in balancing security and usability. A lower threshold increases user friendliness by allowing a wider range of acceptable answers, including those that may be semantically correct but phrased differently. However, this leniency reduces security by making it easier for incorrect answers to be accepted. Conversely, a higher threshold increases security by requiring the user's answer to closely match the model answer, but this strictness may burden legitimate users by requiring more precise recall or by necessitating additional questions to reach the passing criteria. The threshold therefore functions as a tunable parameter that allows system designers to calibrate the trade-off between authentication strength and user experience according to the risk profile of the deployment environment.

After each question is evaluated, the system marks it as passed or failed based on the combined LLM judgment and cosine-similarity scoring. Once all questions have been processed, the system counts the number of passed questions and compares this count to a predefined passing requirement. If the user meets or exceeds this requirement, authentication succeeds; otherwise, authentication fails. For users who fail but have not exceeded the maximum number of allowed attempts, the system resets the question-generation counter and initiates a new round of question creation to let the user retry. Through this structured pipeline — document segmentation, LLM-based question generation, dual-mode answer evaluation, and threshold-based scoring — the system provides a flexible, adaptive, and security-aware authentication mechanism grounded in personalized document understanding.

[1] Python's SentenceTransformer is used to embed the user's answers and the LLM-generated answers into vectors.

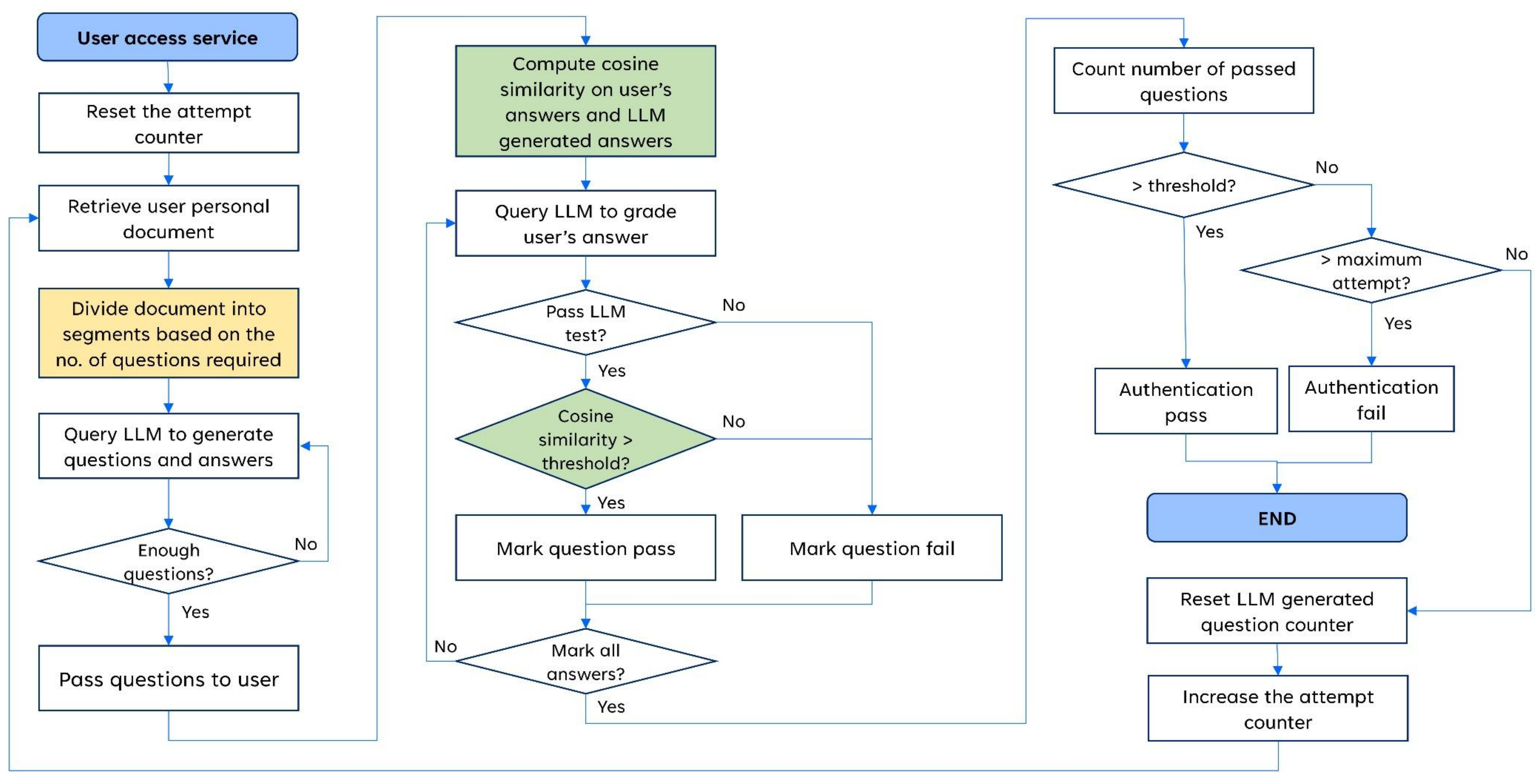

Figure 3. Flow chart of the LLM-assisted User Authentication System

### 3.4 Advantages

The LLM-assisted user authentication method introduces a new variant of the "what-you-know" authentication factor. Its key advantage over traditional approaches such as passwords is that it preserves security assurance without requiring exact, word-for-word matches. Users can respond naturally — as if in conversation — making the process more inclusive and less stressful. This turns authentication into a more user-friendly experience. Moreover, instead of enforcing strict string matching, the system can adapt dynamically by increasing the number of security questions when needed to maintain the desired level of security.

### 3.5 Key Innovations

Figure 4 presents the IP map illustrating how LLM-assisted user authentication diverges from existing authentication methods, and Table 5 summarizes the key attribute differences. Although LLMs have been applied to cybersecurity tasks such as insecure-code detection and cyber-threat intelligence analysis, this work appears to be the first to use their semantic-reasoning capabilities directly for user authentication. It also introduces techniques that address two major limitations of LLMs in this setting: positional bias when generating questions from documents and the untunable, black-box nature of their decision-making.

The proposed method offers two core innovations: (1) balanced security-question generation through document partitioning, and (2) a hybrid scoring mechanism that combines LLM semantic evaluation with statistical similarity measures. To counter the tendency of LLMs to over-sample from the beginning and end of a document, each document is divided into equal segments, and the model is instructed to generate an equal number of questions from each segment. This produces a more uniform question distribution. The hybrid scoring mechanism further enhances robustness by comparing the user's answer with the LLM-generated reference answer using a statistical similarity metric, introducing a tunable parameter that lets practitioners adjust the balance between usability and security.

LLM-assisted authentication reframes the traditional "what-you-know" factor by replacing rigid, exact-match verification with flexible, semantically grounded evaluation. Unlike passwords, preset security questions, or soft-token systems — which demand precise string matching and impose significant cognitive load — this approach allows users to respond naturally in their own words. The conversational interaction improves inclusiveness and reduces stress, particularly for users who struggle with memorization or exact recall.

Compared with biometrics, which provide strong security but rely on immutable traits and are difficult to reset, LLM-assisted authentication remains fully resettable and adaptable. It can also draw on operational data — such as purchase histories or personal documents — to generate personalized challenge questions, enabling richer, context-aware verification while preserving privacy through controlled, system-side data handling.

This invention also differs substantially from prior LLM-related authentication research. Earlier work uses LLMs to assist authentication workflows — for example, composing multi-factor prompts or supporting helper networks — but does not treat the LLM itself as an authentication factor. These systems still rely on exact-match credentials and therefore do not address the core usability limitations of rigid password or question-answer matching. In contrast, this method positions the LLM as the evaluator of semantic correctness, enabling acceptance of answers that are not exact matches but are meaningfully correct.

Two technical contributions reinforce its novelty: document segmentation to mitigate positional bias, and a hybrid scoring mechanism that blends LLM semantic evaluation with cosine-similarity metrics to provide fine-grained control over the usability–security tradeoff. Together, these contributions establish LLM-assisted authentication as a novel, more human-centric, and more adaptable alternative to existing authentication methods and prior LLM-based approaches.

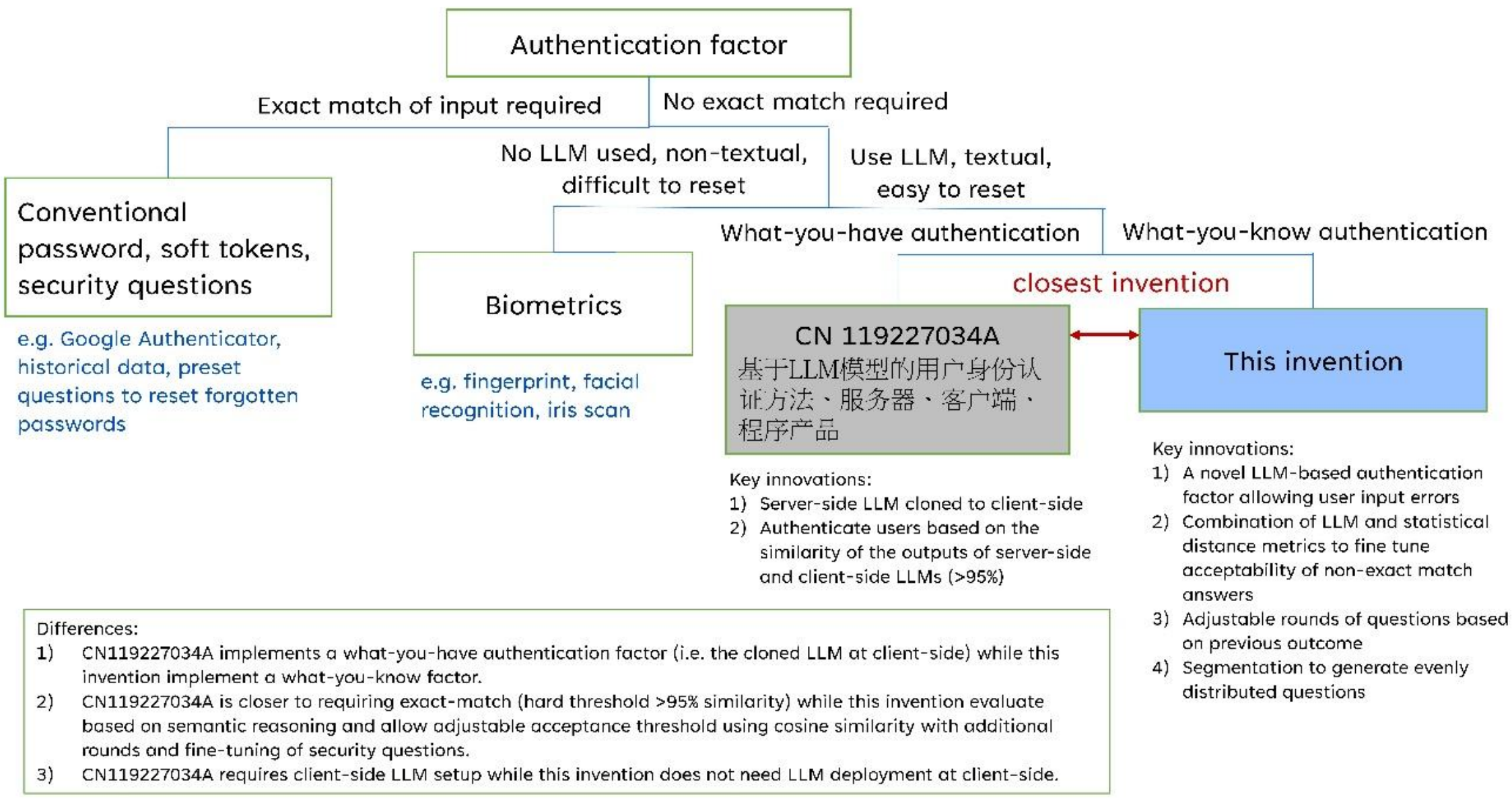

Figure 4. IP Map for LLM-assisted User Authentication

| | Password | Preset question set | Biometrics | This invention |
|---|---|---|---|---|
| Human friendliness for memorization | Low | Medium | High | High |
| Exact match required | Yes | Yes | No | No |
| Leveraging operational data (e.g. purchase records) possible | No | No | No | Yes |
| Adjustable level of security | No | No | No | Yes |
| Adaptive challenge possible | No | No | No | Yes |
| Can be reset | Yes | Yes | No | Yes |

Table 1. Comparison between Different Authentication Methods

## 4. Experiments and Results (LLM-assisted Authentication)

### 4.1 Experiment Setup

Llama-3.3-70B on Hugging Face is used as the LLM for this experiment. A dataset of 1,000 security questions and corresponding model answers is prepared (Appendix A).[2] For each question, two types of modified answers are generated:

- **Correct but non-exact answers**, created by adding or removing non-essential words that do not change the meaning.
- **Incorrect answers**, representing wrong responses.

To quantify similarity, each model answer and its modified counterpart are converted into sentence-embedding vectors, and the cosine similarity between the two vectors is computed.

Each security question is then paired with one modified answer — either a correct non-exact answer or an incorrect one — and the LLM is asked to judge whether the answer should be accepted. The same security question can be used multiple times depending on the number of modified answers it has. The prompt instructs the model to evaluate meaning rather than exact wording:

> "A security question and answer pair is used to authenticate users. A correct answer grants the user access to a protected resource. You are given a security question, its model answer stored in the system, and a user-provided answer. Your task is to determine whether the user's answer should be accepted. Do not grade based on exact wording. Evaluate the meaning of the answer: if its content aligns with the model answer, the user should be granted access."

[2] The dataset of security questions and answers is available at https://huggingface.co/datasets/emunah/authentication.

The LLM's judgment is then combined with the cosine similarity score to determine whether the answer is accepted. For each trial, the numbers of false acceptances (incorrect answers accepted) and false rejections (correct non-exact answers rejected) are recorded. The entire process is repeated three times, and the results are averaged to compute the false acceptance rate and false rejection rate.

### 4.2 Results

Figure 5 and Figure 6 show the false rejection rate and false acceptance rate, respectively, for cosine-similarity thresholds ranging from 0.7 to 1.0. When the threshold is set to 1, only exact matches are accepted, resulting in a false acceptance rate of 0. As the threshold decreases, more non-exact but semantically similar answers are accepted, causing the false acceptance rate to rise. This threshold therefore serves as a tunable parameter that allows practitioners to balance user-friendliness against security assurance.

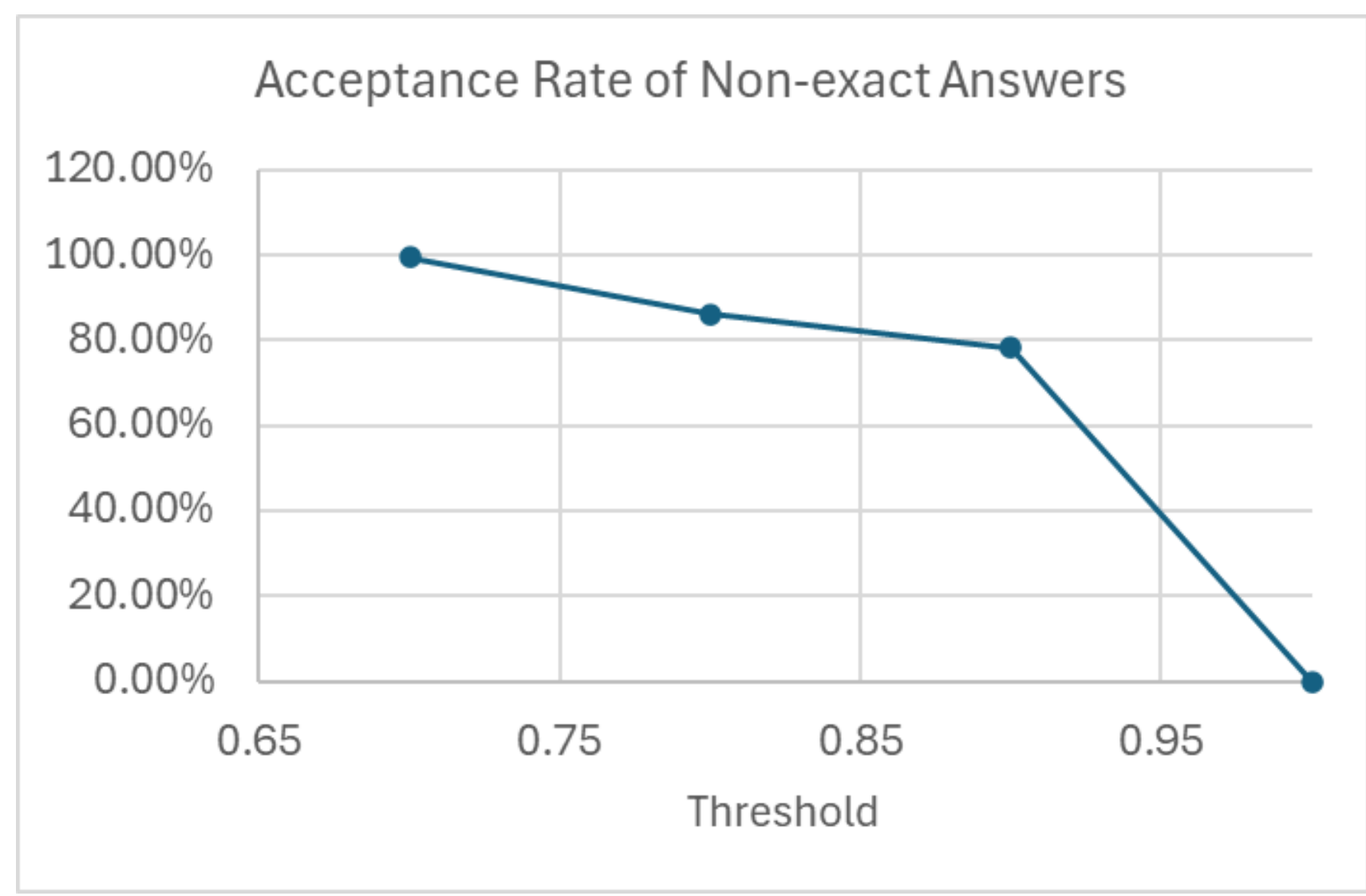

Figure 5. False Rejection Rate of Non-exact Answers

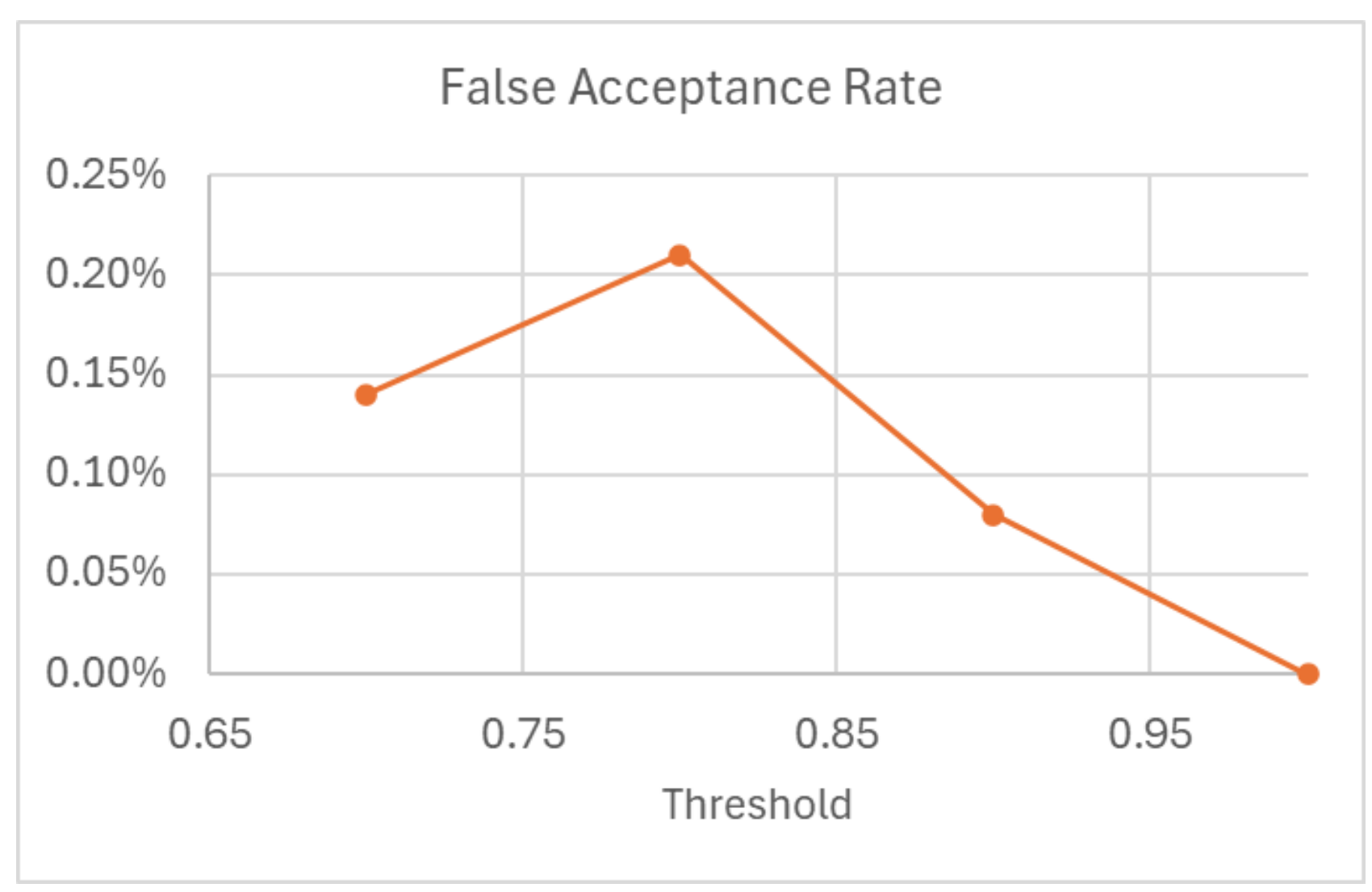

Figure 6. False Acceptance Rate of Wrong Answers

## 5. LLM-Assisted Fraud Detection Using RAG

### 5.1 Fraud Detection Workflow

Figure 7 illustrates the workflow of the proposed LLM-enabled scam-detection system. This design introduces a dynamic, context-driven method for identifying fraudulent messages by integrating LLMs with RAG. Traditional scam-detection approaches, such as classifier-based models, struggle to keep pace with the rapidly evolving tactics used by fraudsters [19]. Unlike these static methods, which rely on fixed rules, keyword matching, or hand-crafted features, the proposed system leverages the semantic reasoning capabilities of LLMs to interpret messages in a human-like and context-aware manner.

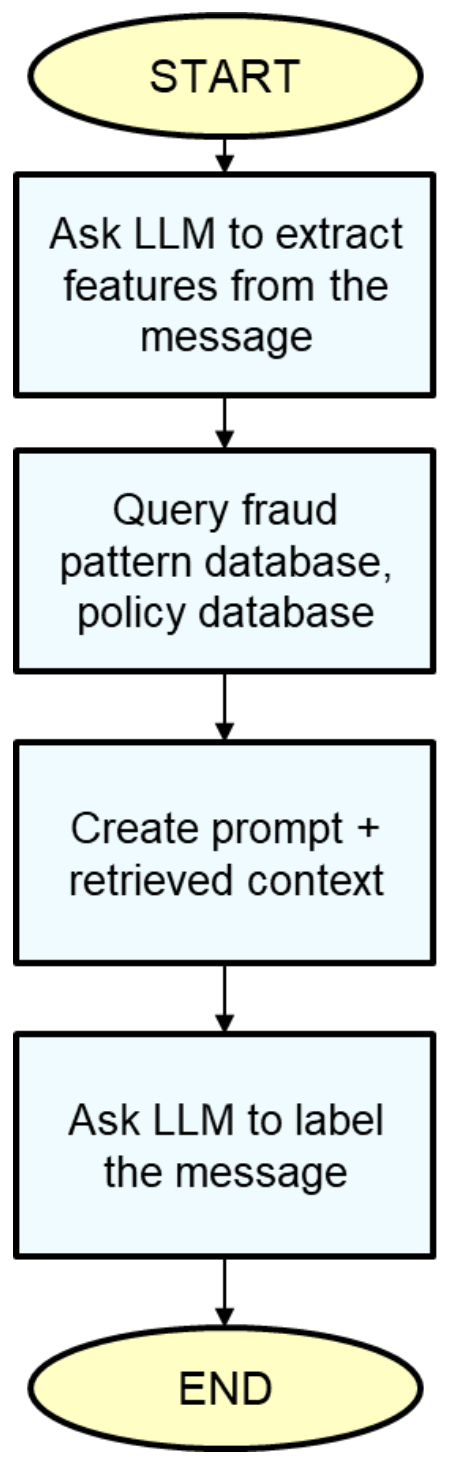

Figure 7. Pipeline for RAG-based LLM Fraud Detection

The process begins with the LLM analysing an incoming message and extracting key features, including intent, tone, urgency, requested actions, referenced entities, and other contextual signals. These extracted attributes are then used to perform targeted retrieval across multiple external knowledge sources, such as verified scam databases, organisational policy documents, and domain-specific reference materials. The retrieved evidence is compiled into a structured prompt that grounds the LLM's reasoning in verified and up-to-date information.

With this enriched context, the LLM conducts a final assessment, producing both a scam-likelihood score and an explanatory rationale. This dual output enhances transparency and facilitates downstream auditing or human review.

A core innovation of the design is its robustness against evolving fraudster behaviour. Because scammers frequently modify their language, structure, and psychological manipulation strategies, static models degrade quickly and require continual retraining. In contrast, the RAG-based architecture adapts naturally: updates occur at the database level rather than within the LLM itself. Newly observed scam patterns can be added to the retrieval corpus, immediately influencing the model's decisions without altering its parameters. This grounding in dynamic evidence also mitigates hallucinations and significantly reduces false-positive rates. Our experiments with ChatGPT-4 show that LLMs tend to over-classify legitimate messages as scams. Incorporating RAG helps anchor the model's judgement in confirmed scam examples, reducing unnecessary misclassifications.

By combining semantic understanding, evidence-based reasoning, and an adaptive retrieval layer, the system provides a scalable, explainable, and resilient framework for fraud detection that surpasses traditional classifier-based approaches and remains effective amid continuously shifting scam tactics.

### 5.2 Preliminary Experiment Results

A dataset of synthetic scam messages is generated based on the dataset in [9] and then combined with a set of legitimate messages. This mixed dataset is used to evaluate ChatGPT-4 by measuring its false-positive rate (legitimate messages incorrectly classified as fraudulent) and its false-negative/recall rate (scam messages incorrectly classified as legitimate).

To assess the performance of the RAG-enhanced detection method, a fresh instance of ChatGPT-4 was instructed to label a collection of confirmed scam messages with specific red-flag attributes, such as the presence of fake links, account-lock threats, urgent language, and similar indicators. This labelled dataset (Appendix B) served as the retrieval corpus for the RAG component. After clearing the model's memory, ChatGPT-4 was tested again on the same mixed dataset, this time with RAG enabled, and its performance was recorded.

Table 2 summarises the results of ChatGPT-4 with and without RAG. As shown, incorporating RAG reduces both false-positive and false-negative rates, demonstrating its effectiveness in improving detection accuracy.

| | LLM without RAG | LLM with RAG |
|---|---|---|
| False negative/recall rate | 7.30% | 5.80% |
| False positive rate | 17.20% | 3.50% |

Table 2. Performance of LLM-based Scam Detection with and without RAG

**Limitations.** To fully evaluate the adaptability of the RAG-enhanced LLM scam-detection system, further experiments using new datasets that reflect emerging and evolving scam tactics would be necessary. However, the scarcity of publicly available, up-to-date information on real-world fraud patterns makes it challenging to construct synthetic datasets that accurately capture these shifting behaviours.

## 6. Conclusion

This work demonstrates that large language models can meaningfully reshape two pillars of cybersecurity, namely, user authentication and fraud detection. By reframing authentication as a semantic evaluation task rather than a rigid string-matching exercise, the proposed LLM-assisted authentication mechanism introduces a new, human-centric variant of the “what-you-know” factor. Document segmentation mitigates positional bias in question generation, while the hybrid scoring method — combining LLM reasoning with cosine-similarity thresholds — provides a tunable balance between usability and security. Experimental results confirm that this approach preserves security guarantees while dramatically improving user friendliness, particularly for individuals who struggle with precise recall.

In parallel, the RAG-based fraud-detection pipeline addresses the limitations of standalone LLM classifiers by grounding their judgments in curated, auditable evidence. This design reduces hallucinations, improves calibration, and enables rapid adaptation to emerging scam tactics without retraining the underlying model. Preliminary experiments show substantial reductions in both false-positive and false-negative rates, underscoring the value of retrieval-anchored reasoning in dynamic threat environments.

Taken together, these contributions highlight a broader shift in cybersecurity: from deterministic, pattern-matching systems toward adaptive, context-aware, and linguistically grounded mechanisms. Rather than replacing existing security controls, LLM-enabled methods augment them with semantic understanding, explainability, and flexibility.

## Appendix A: Samples of the dataset of security questions and answers

Below are sample security questions and answers from the crafted dataset. For the full dataset, please visit https://huggingface.co/emunah/datasets.

```
[
 {
  "question": "Which MTR station did you exit most recently?",
  "answer": "Admiralty",
  "label": true
 },
 {
  "question": "What was the name of your first pet?",
  "answer": "Coco",
  "label": false
 },
 {
  "question": "What is the name of the street you grew up on?",
  "answer": "Des Voeux Road",
  "label": true
 },
 {
  "question": "What is your mother's maiden name?",
  "answer": "Ng",
  "label": true
 },
 {
  "question": "What is your favorite Hong Kong movie?",
  "answer": "The Grandmaster",
  "label": false
 },
 {
  "question": "Which cinema did you last visit?",
  "answer": "Broadway Cinematheque",
  "label": false
 },
 {
  "question": "What was the name of your primary school?",
  "answer": "St. Paul's Co-educational College",
  "label": true
 },
 {
  "question": "What was the make of your first car?",
  "answer": "Mazda 3",
  "label": false
 },
 {
  "question": "What was the name of your primary school?",
  "answer": "La Salle College",
  "label": true
 },
 {
  "question": "Which cinema did you last visit?",
  "answer": "UA iSQUARE",
  "label": true
 },
 {
  "question": "What is your best friend's nickname?",
  "answer": "Fatty",
  "label": false
 },
```

```
 {
  "question": "What is the name of the street you grew up on?",
  "answer": "Canton Road",
  "label": true
 },
 {
  "question": "What is your favorite local restaurant?",
  "answer": "Tsui Wah",
  "label": false
 },
]
```

## Appendix B: Samples of the scam message dataset

Below are sample security questions and answers from the crafted dataset. For the full dataset, please visit https://huggingface.co/emunah/datasets.

```
[
 {
  "id": 1,
  "message": "Your profile security score dropped after multiple failed logins. Re-enable 2-step verification at [FAKE LINK] and confirm recovery email [MASKED EMAIL]. (Action required)",
  "category": "Account/Bank Phishing",
  "scam_type_subcategory": "credential_harvest",
  "channel": "email",
  "urgency_level": "high",
  "red_flags": [
   "account lock threat",
   "urgent language"
  ]
 },
 {
  "id": 3,
  "message": "Unauthorized charge detected. Dispute at [FAKE LINK]. (No reply)",
  "category": "Account/Bank Phishing",
  "scam_type_subcategory": "account_lock_phish",
  "channel": "sms",
  "urgency_level": "high",
  "red_flags": [
   "fake link",
   "account lock threat"
  ]
 },
 {
  "id": 4,
  "message": "[Case: CASE-H5344TF] We prevented a sign-in from a new device in Singapore. To keep your account active, complete verification immediately: [FAKE LINK]. If ignored, limited access may apply. (Action required)",
  "category": "Account/Bank Phishing",
  "scam_type_subcategory": "account_lock_phish",
  "channel": "email",
  "urgency_level": "low",
  "red_flags": [
   "urgent language",
   "spoofed security alert"
  ]
 },
 {
  "id": 5,
  "message": "Security check needed to avoid suspension. Review at [FAKE LINK]. (Auto message)",
  "category": "Account/Bank Phishing",
  "scam_type_subcategory": "account_lock_phish",
  "channel": "email",
  "urgency_level": "low",
  "red_flags": [
   "fake link",
   "account lock threat"
  ]
 },
 {
  "id": 7,
  "message": "Alert: a transfer attempt was blocked (Ref REF-QI3OGZ5K). Review recent activity and confirm your identity immediately via [SAFE REVIEW PAGE]. (Auto message)",
```

```
  "category": "Account/Bank Phishing",
  "scam_type_subcategory": "fraud_alert_spoof",
  "channel": "in-app",
  "urgency_level": "medium",
  "red_flags": [
   "fake link",
   "urgent language"
  ]
 },
 {
  "id": 9,
  "message": "Unauthorized charge detected. Dispute at [FAKE LINK]. (System alert)",
  "category": "Account/Bank Phishing",
  "scam_type_subcategory": "fraud_alert_spoof",
  "channel": "email",
  "urgency_level": "medium",
  "red_flags": [
   "urgent language",
   "account lock threat"
  ]
 },
 {
  "id": 11,
  "message": "Password reset requested. If you did not request this, cancel here: [FAKE LINK]. (No reply)",
  "category": "Account/Bank Phishing",
  "scam_type_subcategory": "account_lock_phish",
  "channel": "sms",
  "urgency_level": "medium",
  "red_flags": [
   "account lock threat",
   "spoofed security alert"
  ]
 },
{
  "id": 13,
  "message": "Alert: a transfer attempt was blocked (Ref REF-N659O2V2). Review recent activity and confirm your
        identity immediately via [SAFE REVIEW PAGE]. (No reply)",
  "category": "Account/Bank Phishing",
  "scam_type_subcategory": "account_lock_phish",
  "channel": "sms",
  "urgency_level": "medium",
  "red_flags": [
   "urgent language",
   "spoofed security alert"
  ]
 },
 {
  "id": 14,
  "message": "Your account has been locked due to unusual activity. Verify at [FAKE LINK] to restore access.",
  "category": "Account/Bank Phishing",
  "scam_type_subcategory": "account_lock_phish",
  "channel": "in-app",
  "urgency_level": "high",
  "red_flags": [
   "fake link",
   "urgent language"
  ]
 }
]
```